\documentclass{article}
\usepackage{spconf,amsmath,graphicx}
\usepackage{tabularray}
\usepackage{algpseudocode}
\usepackage{algorithm}
\usepackage{xcolor}
\usepackage{hyperref}
\usepackage{url}
\usepackage{balance}
\usepackage{multirow}
\usepackage{booktabs}

\title{Learning Arousal-Valence Representation from Categorical Emotion Labels of Speech
}
\name{Enting Zhou$^{1,2}$, You Zhang$^3$, Zhiyao Duan$^3$\thanks{This work is supported in part by a New York State Center of Excellence in Data Science award, and synergistic activities funded by National Science Foundation grant DGE-1922591.}}
\address{$^1$Department of Computer Science, University of Rochester, NY, USA\\
$^2$Department of Electrical and Computer Engineering, University of California, San Diego, CA, USA \\
$^3$Department of Electrical and Computer Engineering, University of Rochester, NY, USA}
\begin{document}
\ninept
\maketitle
\begin{abstract}
Dimensional representations of speech emotions such as the arousal-valence (AV) representation provide a continuous and fine-grained description and control than their categorical counterparts. They have wide applications in tasks such as dynamic emotion understanding and expressive text-to-speech synthesis. Existing methods that predict the dimensional emotion representation from speech cast it as a supervised regression task. These methods face data scarcity issues, as dimensional annotations are much harder to acquire than categorical labels. In this work, we propose to learn the AV representation from categorical emotion labels of speech. We start by learning a rich and emotion-relevant high-dimensional speech feature representation using self-supervised pre-training and emotion classification fine-tuning. This representation is then mapped to the 2D AV space according to psychological findings through anchored dimensionality reduction. 
Experiments show that our method achieves a Concordance Correlation Coefficient (CCC) performance comparable to state-of-the-art supervised regression methods on IEMOCAP without leveraging ground-truth AV annotations during training. This validates our proposed approach on AV prediction. Furthermore, visualization of AV predictions on MEAD and EmoDB datasets shows the interpretability of the learned AV representations. 

\end{abstract}
\begin{keywords}
speech emotion recognition, emotion representation learning, categorical emotions, arousal-valence representation, self-supervised learning
\end{keywords}
\section{Introduction}
\label{sec:intro}

Human interpretable emotion representations are crucial to emotion analysis, and there are generally two kinds of emotion representations: categorical (e.g., angry, happy, sad)~\cite{Ekman1979-ct} and dimensional (e.g., arousal-valence (AV) representation)~\cite{Russell1977-pp}. While categorical representations are easy to understand and annotate, they ignore the nuanced variations within the same category and cannot smoothly connect different categories. Dimensional representations mitigate these shortcomings by mapping emotions onto a multi-dimensional space, offering a continuous and fine-grained description.

Considering that speech serves as a fundamental medium for human interaction and emotional expression, extensive research has been conducted in the domain of speech emotion recognition (SER). Led by the two primary paradigms in emotion representations, there are also two primary kinds of emotion recognition tasks: one classifies emotions into predefined categories, and the other performs regression on dimensional annotations. 
For classification, recent SER systems \cite{Zhao2019-ii, Etienne2018-js, Wang2021-od} employ deep learning to automatically extract speech features and have made substantial improvements on classification accuracy. However, the classification setup is limited to recognizing the predefined discrete emotions.

Recognizing dimensional emotion in speech allows for a nuanced understanding of emotional states, enhancing tasks such as fine-grained emotion-controlled text-to-speech generation~\cite{Sivaprasad2021-ct, Rabiee2019-pz, Cai2021-qi} and emotion analysis~\cite{Mantyla2016-rd}. As a result, there has been considerable research aimed at SER for dimensional emotion prediction. Early dimensional emotion models develop individual models for each dimension, such as distinct valence and arousal estimators~\cite{Wollmer2008-jh}. Such methods ignore the inter-correlation between different emotion dimensions~\cite {Russell1977-pp, Russell1978-ex}. More recent methods~\cite{Atmaja2020-ja, Srinivasan2022-yn, Li2021-dc, Wagner2022-ft} have predominantly adopted Parthasarathy and Busso's framework~\cite{Parthasarathy2017-ws}, which considers the interaction between different dimensions and formulates the problem as a multi-dimensional regression problem. 
The training objective is to minimize the distance or to maximize the correlation between the predicted values and the ground-truth human annotations. However, this setup requires labor-intensive annotations that are much more expensive to obtain compared to categorical labels. Furthermore, such annotations are more susceptible to biases of annotators~\cite{Lotfian2019-qk}. Up to date, there have only been a few emotional speech corpora with such dimensional labels. 

This paper proposes a new approach to learning the AV dimensional representation of speech emotions. Instead of casting it as a supervised regression task as most existing methods do, we propose to learn the representation with self-supervised pretraining and categorical emotion classification finetuning on speech utterances, followed by an anchored dimensionality reduction to the AV space using psychologically informed anchors. 
The rich and emotion-relevant representation in the first step carries sufficient information that can not only distinguish different emotional categories but also demonstrate subtle variations, such as expression styles within the same emotion category. The anchored dimensionality reduction in the second step, mapping latent representation to the AV space, ensures the correct positioning of different emotion categories in the arousal-valence space.
This approach requires only categorical emotion labels in the training data but not dimensional labels, which are much more scarce and difficult to obtain than categorical labels. %
Experiments show that our method achieves comparable Concordance Correlation Coefficient (CCC) performance compared to state-of-the-art supervised regression methods on IEMOCAP. Further tests on EmoDB and MEAD underscore our method's generalizability on diverse speech corpora and show the interpretability of the learned 2D AV representation. Our code is available online\footnote{\url{https://github.com/ETZET/SpeechEmotionAVLearning}}.

\section{Methods}
\label{sec:format}

\begin{figure}[t]
\centering
\includegraphics[width=0.49\textwidth]{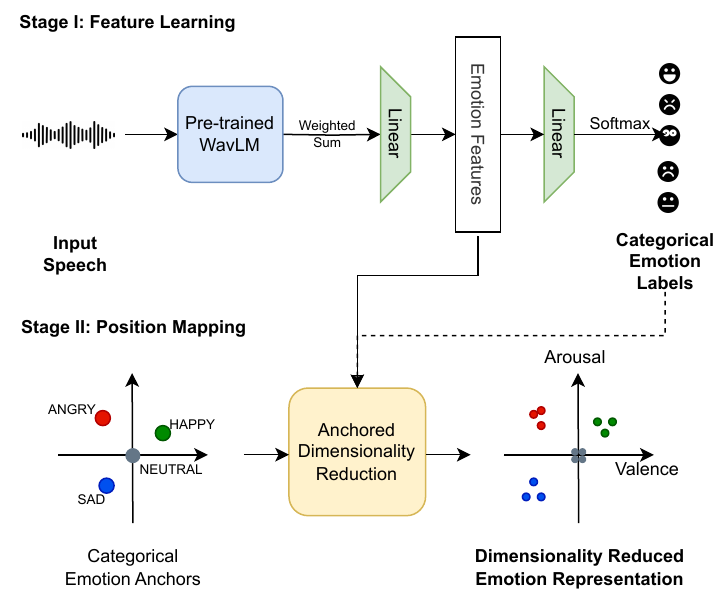}
\caption{Framework of our proposed two-stage arousal-valence learning method. The categorical emotion labels are required during training but optional during inference.}
\label{fig:pipeline}
\end{figure}

In this section, we outline our methodology for arousal-valence (AV) representation learning, which employs a self-supervised front-end and categorical emotion labels of speech utterances to train. The learning process contains two stages as illustrated in Fig.~\ref{fig:pipeline}. In the first stage, we build off from a pre-trained self-supervised speech model (e.g., WavLM \cite{Chen2022-jt}) and train an emotion classification head to derive a high-dimensional latent feature representation. We believe that this feature, derived from SSL pretraining and refined for speech emotion classification, carries rich and emotion-relevant information that can differentiate subtleties across and within emotion categories. 
In the second stage, we introduce an anchored dimensionality reduction technique to project this high-dimensional feature into the 2D AV space, with different emotion categories mapped to their respective anchors informed by psychological studies. 
The whole process only requires categorical emotion labels in the training data. During inference, AV values can be predicted from new speech utterances. 

\subsection{Utterance-level emotion feature extraction}

Self-supervised speech representations are trained using the pretext task of predicting speech tokens based on the surrounding context. Besides achieving leading results on downstream emotion classification tasks, these representations retain structured details within speech utterances, including semantics, speaker data, and paralinguistic cues. We posit that such information is crucial for emotion representation learning,  in the absence of arousal-valence supervision. To extract emotionally relevant features from the input speech, we address an emotion classification challenge,


We adopt WavLM Large~\cite{Chen2022-jt} as the self-supervised front-end as it achieves state-of-the-art results and marks a significant improvement on various benchmark speech downstream tasks~\cite{Yang2021-in} over other self-supervised front-ends, in particularly in emotion classification on several benchmark emotion speech dataset \cite{atmaja2022evaluating}.
To build off from WavLM, we add a weighted projection head for categorical emotion classification. Following the design of~\cite{Chen2022-jt} and~\cite{Yang2021-in}, we compute the weighted sum of the twenty-four transformer layers, each with a 1024 dimension, and average pool temporally to derive utterance-level embeddings. Subsequently, the averaged-pool weighted sum output is passed through a linear projection layer to reduce the feature dimension to 100 before reaching the classification head. This feature vector is then forwarded to another linear layer that further reduces its dimensionality to match the number of emotions in the training set, followed by a softmax function for normalization. We optimize the network with cross-entropy loss for multi-class classification. The model's 100-dimensional output from the second last layer serves as the high-dimensional emotion feature, as illustrated in the top part of Fig.~\ref{fig:pipeline}. Note that this feature is a linear mapping of the self-supervised learned features, where the weights are trained for categorical emotion classification. This feature is believed to encapsulate rich emotion-related data in speech utterances for AV representation learning.


\begin{table}
\centering
\caption{The arousal-valence (AV) values as emotion anchors for ten categorical emotions in the position mapping stage. These values are obtained from psychology study~\cite{Russell1977-pp}.}
\label{table:referenceAV}
\begin{tblr}{
  colsep = 0.3em,
  rowsep = 0.0em,
  cells = {c},
  hline{1-2,7} = {-}{},
  vline{4} = {-}{},
}
Emotion  & Valence & Arousal & Emotion    & Valence & Arousal \\
Angry    & -0.51   & 0.59    & Fearful       & -0.64   & 0.60    \\
Boredom  & -0.65   & -0.62   & Frustrated & -0.64   & 0.52    \\
Contempt & -0.80   & 0.20    & Happy      & 0.81    & 0.51    \\
Disgusted  & -0.60   & 0.35    & Sad        & -0.63   & -0.27   \\
Excited  & 0.62    & 0.75    & Surprised  & 0.40    & 0.67    
\end{tblr}
\end{table}

\subsection{Categorical emotion label-guided arousal-valence representation learning}

To predict the arousal and valence values of the speech utterance, we leverage categorical labels to guide the representation learning from the high-dimensional feature space to the arousal-valence (AV) space. Inspired by the inherent connection between categorical and dimensional representation \cite{Hoffmann2012-cr}, We first assign each categorical label with certain AV values as ``anchors'', based on psychology literature~\cite{Russell1977-pp}. These anchors provides a foundational guide that underpin our subsequent data-driven approach to learn AV representation. The reference anchors used in this study are presented in Table \ref{table:referenceAV}. The final predicted AV values of the speech utterances are expected to center around those anchors. Note that the neutral emotion was not covered in their study; hence, we position the neutral emotion at the origin of the arousal-valence space, i.e., setting the AV values for neutral as zeros. Subsequently, we optimize embedded points' position starting from these anchors but retaining structural information of the emotion features. 



We start by setting the embedded points' positions in the arousal-valence space for each speech utterance close to the anchor corresponding to their categorical label, i.e., assigning initial arousal-valence values to the anchor value with a small Gaussian perturbation $\sigma = 0.01$. Subsequently, we fine-tune the positions of the embedded points based on the high-dimensional speech emotion features obtained from Stage I. To encapsulate the structure of these speech emotion features, we construct a weighted $k$-Nearest Neighbor (kNN) graph. The kNN graph will be regarded as a topological representation of the extracted emotion features. We infuse categorical labels to the constructed kNN graph to encourage class separation in the embedding space. These provided labels define a distinct metric space that intersects with the kNN graph. Computationally, this equates to weighting the distances between points by a constant. 
We utilize a stochastic gradient descent approach similar to UMAP~\cite{McInnes2018-ct} to minimize the cross-entropy between high-dimensional features and low-dimensional embeddings, iteratively refining the position of each point through calculations of both attractive and repulsive gradients.

During inference, our method predicts arousal-valence values for new speech samples without retraining, despite its basis in a nonparametric dimensionality reduction algorithm. We can make inferences with or without categorical labels for new samples. In stage II, when categorical labels are available, we initialize the embedded values of new samples proximate to a reference emotion anchor determined by their categorical labels. We then optimize these points using the extracted emotion features, mirroring the training phase. When the categorical labels are not available, we execute a kNN search on the emotion features of the training dataset, using the features of new samples as queries. The resulting embedded values for each new sample are derived from the weighted average of its neighboring embedding values.

\section{Experimental Setup}
\label{sec:pagestyle}

\subsection{Database description}

In this study, we use IEMOCAP~\cite{Busso2008-pr}, MEAD~\cite{Wang2020-km}, EmoDB~\cite{Burkhardt2005-je} to validate the effectiveness and generalization of our approach. These corpora are commonly used for SER tasks. IEMOCAP serves as our primary benchmark for evaluating predicted AV values, with MEAD and EmoDB assessing the versatile capability of our method across diverse corpora, including cross-lingual and emotional context.

\textbf{IEMOCAP}:
The IEMOCAP database contains 12 hours of audiovisual data with speech utterances from improvised affective scenarios and theatrical script performances. Each segment has annotations from at least three individuals for emotional categories and dimensional values. The final label is decided by majority voting, and dimensional values are averaged from all annotations. The corpus features both categorical emotion annotations and dimensional emotion annotations. Ten emotion categories are featured: angry, happy, neutral, sad, disgusted, frustrated, excited, fearful, surprised, and others. If there's no majority vote, the label 'XXX' is used.

\textbf{MEAD}:
The Multi-view Emotional Audio-visual Dataset consists of audio-visual recordings of 48 professional actors expressing various emotions based on predefined scripts. It covers eight emotions, each with three intensities except for neutral, totaling over 30,000 utterances. The spoken content includes emotion-neutral and emotion-related sentences for each emotion. For this study, we focus solely on the audio portion of the MEAD dataset.

\textbf{EmoDB}:
EmoDB is a German emotional speech corpus comprising 10 speakers (5 male, 5 female) articulating daily sentences with seven emotions: anger, boredom, disgusted, fearful, happiness, neutral, and sad. For this study, we use this corpus to evaluate our methods’ performance in a non-English speaking speech dataset. 

\subsection{Implementation details}

\textbf{Training setup}: The emotion feature extractor is individually fine-tuned for each corpus. Each model is trained for 50 epochs with a batch size of 8, utilizing the Adam optimizer with a learning rate of 1e-4. We use a speaker-independent train-test split for unbiased assessment. For anchored dimensionality reduction, the minimum distance and number of nearest neighbors are set to 0.1 and 20. The points optimization process learning rate is kept low at 1e-2 to maintain guidance initialization.

\textbf{Evaluation metrics}:
We compute the agreement between our predicted arousal-valence value with the human-annotated value using the Concordance correlation coefficient (CCC) \cite{Lin1989-nk} by Eq. \eqref{CCC metric},
\begin{equation}
\label{CCC metric}
    \rho_c = \frac{2\rho_{xy}\sigma_x\sigma_y}{\sigma_x^2 + \sigma_y^2 + (\mu_x - \mu_y)^2},
\end{equation}
where $\rho_{xy}$ is the Pearson correlation coefficient between $x$ and $y$, $\sigma$ is the standard deviation, and $\mu$ is the mean value. This metric offers a distinct advantage over simple Pearson correlation as it can effectively handle annotations with different scales and ranges. We additionally include Mean Absolute Error (MAE) as an additional metric to evaluate the predicted AV representation comprehensively.

\section{Results}
\label{sec:results}


\subsection{Comparison with human annotations}


We compare our predicted arousal-valence (AV) values with the human annotations in the IEMOCAP dataset. For a fair comparison with prior works, we evaluate methods on IEMOCAP with three configurations of emotion subsets: 1) 4 emotions (angry, happy, neutral, sad), 2) 5 emotions (angry, happy, neutral, sad, disgusted), and 3) 9 emotions (all excluding `others' and `XXX').   

In Table \ref{table:CCCPerformance}, we showcase the performance of various methods on the IEMOCAP dataset using CCC and MAE for AV prediction. We compare our methods to recent state-of-the-art regression methods. While our methods fall short in the 4 and 5 emotion subsets, they exhibit competitive performance in the 9 emotion subset. We anticipate that recent supervised regression methods would excel since they are direct learning from ground truth annotations using intricate models and leveraging auxiliary data like categorical labels and textual information. Nonetheless, our model achieves comparable performance in the challenging context of 9 emotion subsets. This underscores the validity of our proposed approach for AV prediction without relying on dimensional annotations during training.

Given the intricate architecture and varying input modalities in recent studies, we established a baseline regression model for a fairer comparison. This model learns to predict arousal-valence values directly from the ground-truth human annotations. This model replicates our stage I method's architecture but replaces the final classification head with two independent linear layers for valence and arousal prediction. The optimization employs the Concordance Correlation Coefficient Loss as employed in \cite{Wagner2022-ft, Srinivasan2022-yn, Atmaja2021-cc}.  We individually trained the regression baseline for each IEMOCAP setup, with results detailed in Table \ref{table:CCCPerformance}. Compared to the regression baseline we constructed, our method consistently outperforms. In the setting where 4 emotions or 5 emotions are involved in training, we achieve CCC values of 0.529 and 0.545 for valence, respectively, and arousal values around 0.63, outperforming the baseline method. In the 9-emotions setting, our method significantly surpasses the baseline in valence while retaining a comparable arousal value. Using only categorical labels for supervision during training, our model consistently performs across all scenarios, highlighting its efficacy in AV predictions.

Driven by scenarios where categorical labels are available during inference, such as the transformation of a categorically annotated affective speech corpus into a dimensionally annotated one, we assess a version of our method using ground truth categorical labels during inference (denoted as "Ours w/ CL"). This significantly enhances the CCC performance for the valence dimension while maintaining a similar performance in arousal. Such findings validate our method's capability to improve AV predictions when supplemented with auxiliary information during testing.

In the 9 emotions setting, we experimented with two reduced versions of our methods, one without the use of emotion anchors and one without supervision from the categorical labels in training. The former obtained an embedding that misaligned with the arousal and valence axis, achieving close to zero CCC. The latter achieves slightly inferior CCC and MAE performance in both dimensions. 

\begin{table}[]
\centering
\caption{Performance comparison on IEMOCAP among regression methods and our approaches using CCC and MAE metrics. ``Ours'' denotes our proposed method predicting AV values from speech, while ``Ours w/ CL'' denotes an extension of our proposed method when the ground-truth categorical labels (CL) are leveraged.}
\label{table:CCCPerformance}
\begin{tblr}{
  cells = {c},
  cell{1}{1} = {r=2}{},
  cell{1}{2} = {r=2}{},
  cell{1}{3} = {c=2}{},
  cell{1}{5} = {c=2}{},
  hline{1,7,11,16} = {-}{},
  hline{3} = {1-6}{},
  rowsep = 0.00em,
  colsep = 0.3em,
  row{6} = {fg=gray!80},  
  row{10} = {fg=gray!80},  
  row{15} = {fg=gray!80}  
}
\#Emotions  & Methods           & Valence        &                  & Arousal        &                  \\
            &                   & CCC $\uparrow$ & MAE $\downarrow$ & CCC $\uparrow$ & MAE $\downarrow$ \\
4           & M. Trnka et al.   & 0.631          & 0.573            & 0.750          & 0.407            \\
4           & Baseline          & 0.443          & 0.447            & 0.614          & 0.314            \\
4           & Ours       & 0.529          & 0.318            & 0.632          & 0.273            \\
4           & Ours w/ CL        & 0.693          & 0.292            & 0.638          & 0.270            \\
5           & Sharma et al.     & 0.660          & -                & 0.717          & -                \\
5           & Baseline          & 0.459          & 0.426            & 0.618          & 0.308            \\
5           & Ours       & 0.545          & 0.325            & 0.630          & 0.264            \\
5           & Ours~w/ CL        & 0.678          & 0.295            & 0.619          & 0.291            \\
9  + others & Atmaja and Akagi. & 0.418          & -                & 0.571          & -                \\
9  + others & Srinivasan et al. & 0.582          & -                & 0.667          & -                \\
9           & Baseline          & 0.317          & 0.490            & 0.718          & 0.214            \\
9           & Ours      & 0.566          & 0.303            & 0.672          & 0.242            \\
9           & Ours~w/ CL        & 0.674          & 0.274            & 0.679          & 0.247            
\end{tblr}
\end{table}

\begin{figure}[t]

\begin{minipage}[t]{0.5\textwidth}
  \centering
  \centerline{\includegraphics[width=0.98\textwidth]{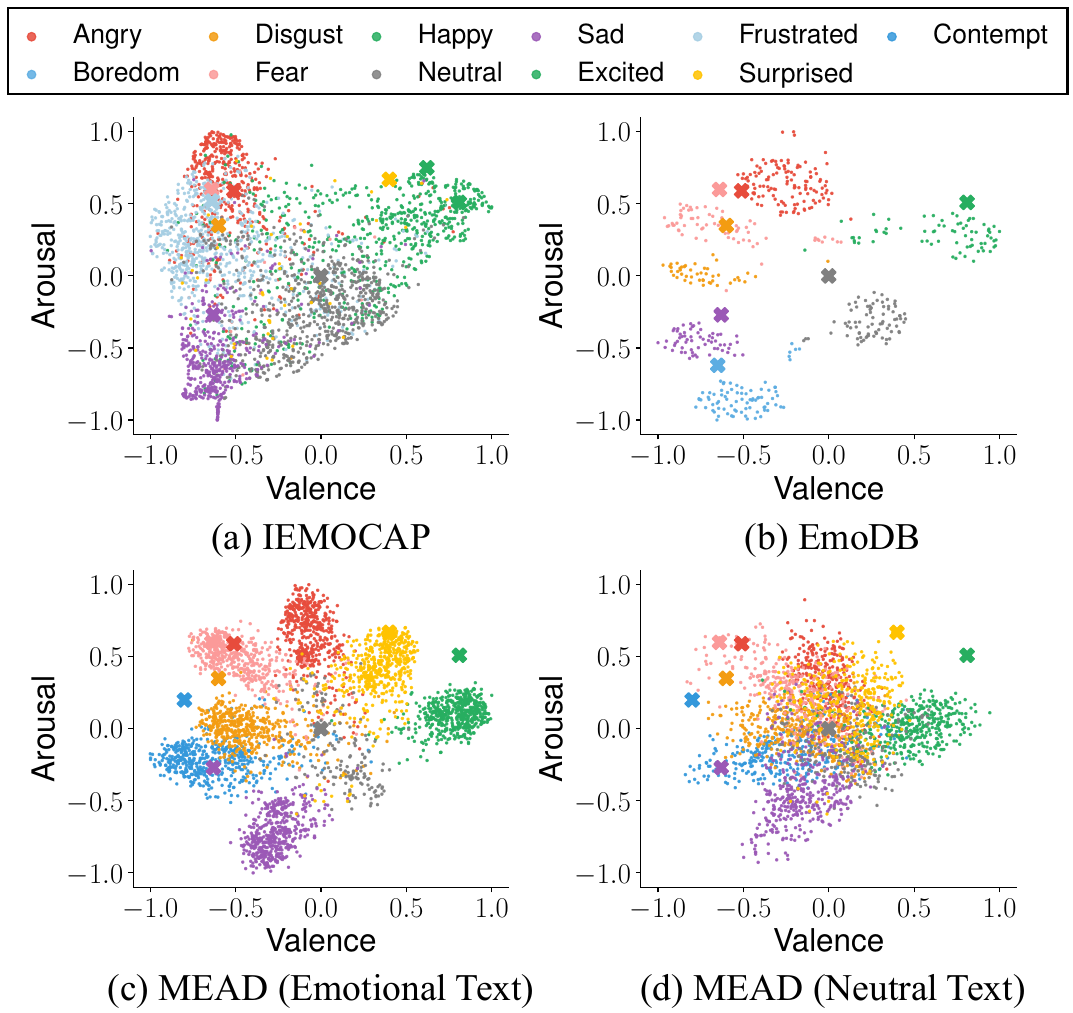}}
\end{minipage}

\caption{Visualization of the Estimated Arousal Valence Values. Emotion anchors are labeled with crosses.}
\label{fig:result}
\end{figure}

\begin{table}
\centering
\caption{Comparison of MAE for valence and arousal predictions/annotations across datasets from psychology emotion anchors.}
\label{table:distanceToAnchor}
\begin{tblr}{
  rowsep=0.00em,
  cells = {c},
  cell{1}{1} = {r=2}{},
  cell{1}{2} = {c=2}{},
  hline{3} = {-}{0.05em},
  hline{1,8} = {-}{0.08em},
}
Dataset                     & MAE     &         \\
                            & Valence & Arousal \\
IEMOCAP (Human Annotations) & 0.253   & 0.402   \\
IEMOCAP (Predicted)         & 0.197   & 0.317   \\
EmoDB                       & 0.141   & 0.221   \\
MEAD (Emotional Text)       & 0.230   & 0.257   \\
MEAD (Neutral Text)         & 0.376   & 0.334   
\end{tblr}
\end{table}



\subsection{Visualization of learned arousal-valence values}


We present the visualization of our learned representation for IEMOCAP, EmoDB, and MEAD in Figure \ref{fig:result}, alongside Table \ref{table:distanceToAnchor} to quantitatively measure the distance between the center of emotion clusters and the guidance anchors on the AV plane. 

We present the visualization of our learned representation on the AV plane for IEMOCAP in Figure \ref{fig:result}(a). Within the embedded AV space, the emotion clusters align closely with our guidance anchors: the neutral cluster is centralized at the origin; clusters representing happiness and excitement exhibit higher valence values than those of anger and sadness; and angry utterances manifest elevated arousal values compared to neutral and sad ones. A low MAE between the cluster center and the guidance anchors further indicates the effectiveness of anchors in dimensionality reduction, leading to a meaningful embedding of speech emotion features within the AV space.

In Figure \ref{fig:result}(b), the embedded AV values in EmoDB foster distinct emotion clusters compared to other corpora we evaluated. This finding suggests that speech utterances in EmoDB exhibit strong prototypical emotion features, which is unsurprising given that EmoDB is an acted corpus with only ten speakers. The result on EmoDB indicates the transfer capability of our approach to infer AV values for speech utterance on a non-English corpus. Figure \ref{fig:result}(c) and \ref{fig:result}(d) present our results on the MEAD dataset. We have an interesting finding when comparing the embedded values of speech acting on the emotional text with those acting on the neutral text. Specifically, the former demonstrates clearly separated clusters, while the latter, across all emotional categories, tends toward the origin. The distance between the cluster centers and anchors also reveals such difference, where the embedded values of the emotional text portion are closer to the centers than those of neutral text. This suggests that the captured emotion features discern semantic nuances in speech utterances, or perhaps actors convey emotions distinctly based on the emotional context. Exploring whether the extracted features encapsulate semantic content warrants future investigation. The visualization shows the learned AV representation aligns with the intrinsic characteristics of emotion distribution of affective speech corpus. The applicability of our method across datasets suggests its potential in harmonizing different emotional speech corpora, aiming for a consistent dimensional representation across multiple datasets.



\vspace{-1pt}

\section{Conclusion}
\label{sec:conclusion}

\vspace{-1pt}

We propose a novel approach to learning arousal-valence (AV) representations for speech from categorical labels, leveraging a combination of self-supervised pretraining and anchor dimensionality reduction. This method outperforms the simple regression baseline in the IEMOCAP dataset and is competitive with current state-of-the-art (SOTA) regression techniques. Visualization of the derived AV values validates the applicability of our methods and shows the interpretability of the learned embeddings. The method can extend to other modalities like visual and textual data and be used for subsequent tasks such as emotion analysis and affective speech synthesis.

\vfill\pagebreak

\bibliographystyle{IEEEbib}
\bibliography{ref}

\end{document}